\documentclass[reprint,
 amsmath,amssymb,
 aps,
 prl,
 superscriptaddress
]{revtex4-1}

\usepackage{graphicx}
\usepackage{dcolumn}
\usepackage{bm}
\usepackage{hyperref}

\begin{document}

\preprint{APS/123-QED}

\title{Mechanical stability of particle-stabilized droplets under micropipette aspiration}

\author{Niveditha Samudrala}
\affiliation{%
Yale University, New Haven, Connecticut 06520, United States of America  
}

\author{Jin Nam}
\affiliation{%
AMOREPACIFIC Co., Gyeonggi-do, Seoul 446-729, South Korea
}

\author{Rapha\"{e}l Sarfati}
\affiliation{%
Yale University, New Haven, Connecticut 06520, United States of America 
}

\author{Robert W. Style}
\affiliation{
ETH Z\"{u}rich, CH - 8093, Switzerland 
}

\author{Eric R. Dufresne}
 \email{eric.dufresne@mat.ethz.ch}
 \affiliation{
ETH Z\"{u}rich, CH - 8093, Switzerland 
}%

\date{\today}
            
\begin{abstract}
We investigate the mechanical behavior of particle-stabilized droplets using micropipette aspiration. 
We observe that droplets stabilized with amphiphilic dumbbell-shaped particles exhibit a two-stage response to increasing suction pressure. 
Droplets first drip, then wrinkle and buckle like an elastic shell.
While particles have a dramatic impact on the mechanism of failure, the mechanical strength of the droplets is only modestly increased.
On the other hand, droplets coated with the molecular surfactant Sodium Dodecyl Sulfate are even weaker than bare droplets.
In all cases, the magnitude of the critical pressure for the onset of  instabilities is set by the fluid surface tension.

\begin{description}

\item[PACS numbers]

\end{description}
\end{abstract}

\pacs{Valid PACS appear here}
\maketitle

\section{I. Introduction}

Emulsions are typically produced by the application of shear to immiscible liquids with the help of a surfactant \cite{Ian}. 
Colloidal particles and amphiphilic molecules are widely used as surfactants.
The fundamental difference is that particles have a much higher binding affinity to the interface \cite{BinksBook}.
The mechanical properties of individual emulsion droplets play an important role in determining the rheology of the emulsion \cite{dropletdeformability,emulsionsweitz} and its stability upon the application of further shear, which can be used to trigger the release of encapsulated fluid \cite{shearrelease}.

The mechanical properties of complex fluid-fluid interfaces are a subject of ongoing investigation \cite{JanGerry}.
Traditionally, surfactant-laden flat interfaces are characterized under compression using a Langmuir-Blodgett trough.
There, the net interfacial tension, $\tau$, is measured as a function of surface coverage.  
In the dilute limit, the net interfacial tension is equal to the bare interface tension, $\gamma$.  
As the coverage increases, the adsorbed components interact and resist compression, lowering the net tension \cite{Butt,Milner}.
At sufficiently high surface coverage, particle-laden interfaces can have solid-like rheology \cite{JanGerry}.
As the net tension on the interface vanishes, particle-laden interfaces become unstable and exhibit buckling or corrugation
 \cite{Aveyard}. 

The mechanics of particle-laden droplets has recently been studied using a variety of approaches \cite{SFMpaper,AFMpaper,VellaTensiometry,Sujit,PitoisEPL,FullerPCCP,FullerLangmuir,Petra}.
The compression of particle-laden droplets, whose radii are in the vicinity of 10 $\mu$m, using Scanning Force Microscopy \cite{SFMpaper} or Atomic Force Microscopy \cite{AFMpaper} revealed substantial deviations from the elastic shell models, with apparent contributions from surface tension.
The aspiration of sessile particle-laden millimetric  droplets observed buckling as the tension of the interface approaches zero \cite{FullerLangmuir, FullerPCCP}.

In this paper, we investigate the mechanics of emulsion droplets using micropipette aspiration. 
While both particle- and molecule-stabilized droplets are stable, they respond very differently to suction pressure.
Droplets stabilized by Sodium Dodecyl Sulfate (SDS) fail at a much lower suction pressure than bare droplets or droplets stabilized with amphiphilic dumbbell particles.
While the SDS-stabilized droplets are sucked into the micropipette like a bare droplet, the particle stabilized droplets undergo a two step failure:
first, fluid is removed from the droplet, and then the particle shell buckles.
In all cases, however, the magnitude of the critical pressure for these instabilities is set by the fluid surface tension.

\section{II. Materials and Methods} 

\paragraph{Colloidal Particles.}  We use sub-micron dumbbell-shaped particles as emulsifiers. 
The dumbbells are bulk synthesized through a two-step seeded emulsion polymerization technique \cite{jingyujacs} and are highly monodisperse.
The largest dimension of the dumbbell particles is 0.5 $\mu$m and the lobe radii are 0.25 $\mu$m. 
However, they have different chemistry on the two lobes; one lobe is polystyrene and the other is a random co-polymer of styrene and trimethoxysilylpropylacrylate. 
As a result, the particles are weakly amphiphilic with anisotropic wetting preferences at alkane-water interface \cite{luciolangmuir}. 

\paragraph{Emulsions.}
We prepare oil-in-water emulsions by taking 1 $m$L n-hexadecane and 9 $m$L of deionized (DI) water in a vial. 
We gently squirt 0.5 $m$L of 10\% (by weight) particles close to the oil-water interface and vortex the vial immediately for a few seconds.
We then use a homogenizer (Ultraturrax T-18) at 10,000 rpm for 60 s to prepare the emulsion. 
The emulsion is then left to age for three days to ensure that the droplet surfaces are jammed with particles.
We repeat this protocol for all the measurements in this paper to control for age and extent of surface coverage.

\paragraph{Micropipettes.} 
We prepare micropipettes by pulling glass capillaries (World Precision Instruments TW100-6) in a micropuller (Sutter Instruments P-1000) and then use a microforge (Narshige Instruments MF-900) to cleanly cut the pipettes radii $R_p$ $\in$ [3,12] $\mu$m.

\paragraph{Aspiration Equipment.} 
We mount the micropipette onto a pipette holder connected to a hydraulic pressure system and a micromanipulator (Narshige MMO-203).
We can regulate the pressure of the hydraulic system from 0 to 30 kPa below atmosphere with 1 Pa resolution. 
The suction pressure $\Delta P$, which is the difference between the pressure inside the pipette $P^{(pip)}$ and the atmospheric pressure $P^{(atm)}$, is read out to an accuracy of 0.14 kPa using a pressure transducer (Validyne Engineering P61D-38S).
The micromanipulator allows us to precisely control the x-, y-, and z-position of the pipette.
The micromanipulator is mounted on a larger 3-axis micrometer stage that allows us to position the entire apparatus appropriately for microscopy.

\paragraph{Aspiration Protocol.} 
We resuspend the buoyant emulsion droplets in a DI water sample chamber for aspiration. 
The sample is zeroed at atmospheric pressure $P^{(atm)}$. 
An emulsion droplet is then aspirated with the micropipette and submerged into DI water to create a uniform environment around the droplet. 
We then increase the suction pressure using a syringe pump (Harvard Apparatus 703006) quasi-statically to control for viscosity effects. 

\paragraph{Imaging.}
We track the structural deformation of the droplet in real time using an inverted optical microscope (Nikon Eclipse TE2000-V).  
We illuminate the sample with a low-N.A. condenser and image using a 40x (N.A. 0.60) air objective.
We acquire images at 1 frame per second for the duration of the experiment using a camera (Thorlabs DC3240M) interfaced with MATLAB.

\paragraph{Analysis.}
We measure the geometrical parameters, droplet radius $R_d$ and pipette radius $R_p$, using ImageJ. 
We analyze the images using a combination of ImageJ and MATLAB.

\section{III. Results and Discussion}

We study the deformation of particle-laden droplets using micropipette aspiration. 
Aspiration has been extensively used in biology to investigate the viscoelastic properties of single cells \cite{studMPApaper} and cell aggregates \cite{Francoise}, stiffness of membranes \cite{Aoki,MPAMembrane}  and most recently, mechanics of emulsion droplets stabilized by bacteria \cite{MPAbacteria}. 
The technique allows direct measurement of the mechanical properties of the interface by applying stress locally on the droplets.
The stress applied on the droplets can be tuned by changing the differential pressure between the pipette and the ambient environment. 
Aspiration combined with microscopy enables us to directly visualize the morphological response of droplets in real time and is particularly well-suited for measuring micrometer-scale perturbations.

Gentle aspiration pressures, about 2 kPa, capture droplets at the tip of the micropipette without any visible deformation.
Time-lapse imaging over a period of 120~s reveals no rearrangement of the colloidal particles on the droplet surface (\textit{Supplemental Movie 1}) indicating that the particles have formed a system-spanning solid network on the droplet. 

\begin{figure}
\includegraphics[width=0.48 \textwidth]{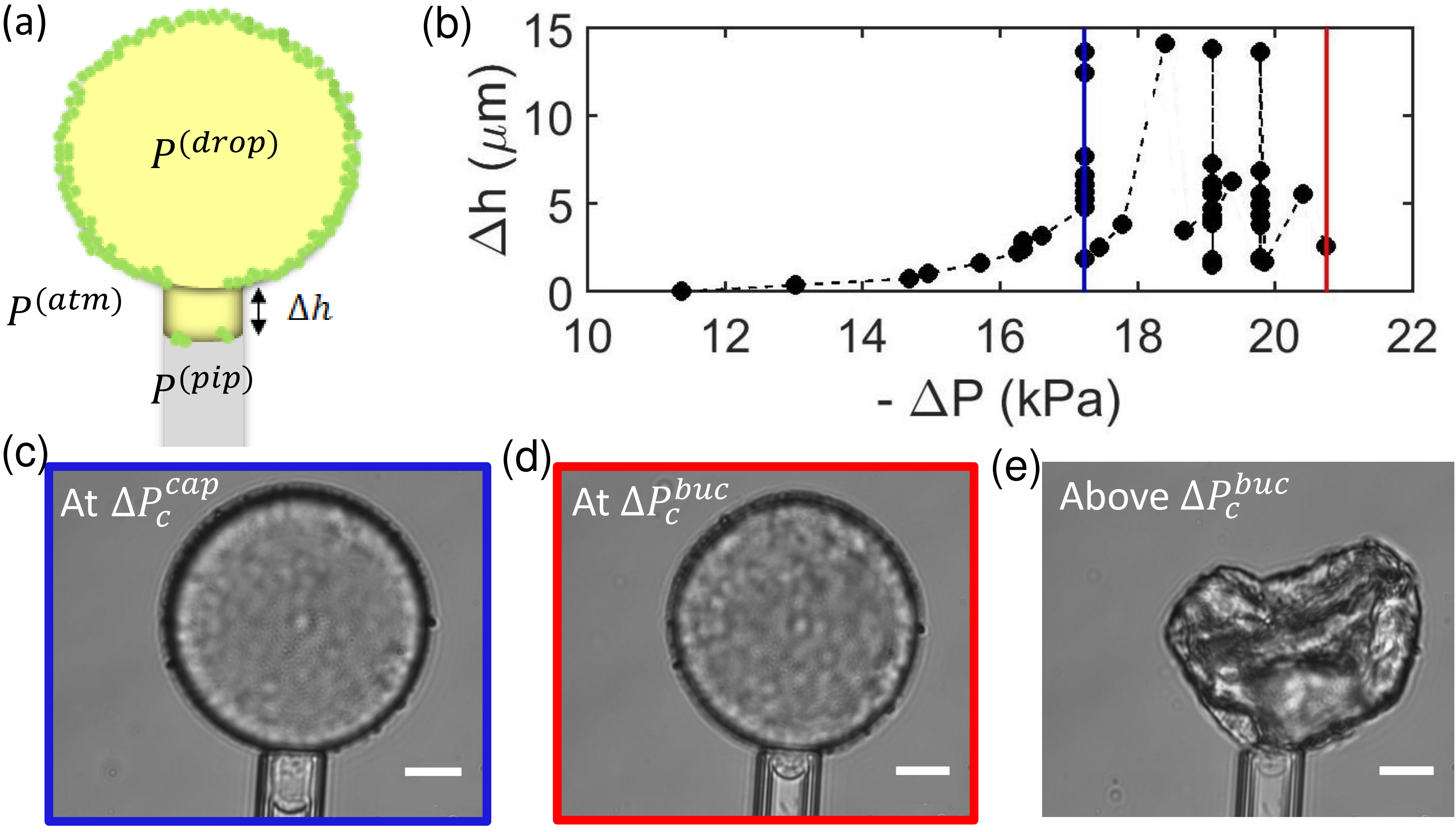}
\caption{
\textit{Micropipette aspiration of particle-laden droplets.} 
 \textbf{(a)} Schematic of the experiment.  
\textbf{(b)} Tongue length, $\Delta h$,  versus suction pressure, $\Delta P=P^{(pip)}-P^{(atm)}$ , which is increased quasi-statically.
The dashed line connects data points continuously leading up to pinch off. 
We identify the onset of capillary instability as the point where $\Delta h$ grows despite holding the pressure constant, $\Delta P^{cap}_{c}$ \textit{(blue line)}.  
The droplet eventually buckles at critical pressure $\Delta P^{buc}_{c}$ \textit{(red line)}.
\textbf{(c-e)} Images of the droplet \textbf{(c)} at the capillary instability, \textbf{(d)} at the buckling instability, and \textbf{(e)} above the buckling threshold.
The scale bar in all images is 15 $\mu$m.
}
\label{Fig1} 
\end{figure}

Upon increasing suction, we observe that a small portion of the particle-laden droplet, or \emph{tongue}, is pulled into the pipette.
We increase the suction pressure quasi-statically and measure the change in the length of the tongue,  $\Delta h$. 
Beyond a pressure threshold, we observe that the tongue becomes unstable and continues to grow even when held at constant suction (Figure \ref{Fig1} (b)).  
We call this a \emph{capillary instability} and the corresponding pressure the \textit{critical pressure} given by $\Delta P^{cap}_{c}$ (blue line in Figure \ref{Fig1} (b)). 
The tongue then grows steadily until it pinches off and then re-establishes a new stable tongue length (\textit{Supplemental Movie 2}).
In contrast, bare droplets show a catastrophic capillary instability: 
once the tongue becomes unstable, the bare droplets are entirely sucked into the pipette (\textit{Supplemental Movie 3}).
Thus, the  particles prevent the droplet from being completely sucked into the pipette after the onset of capillary instability.

\begin{figure}
\includegraphics[width=0.45 \textwidth]{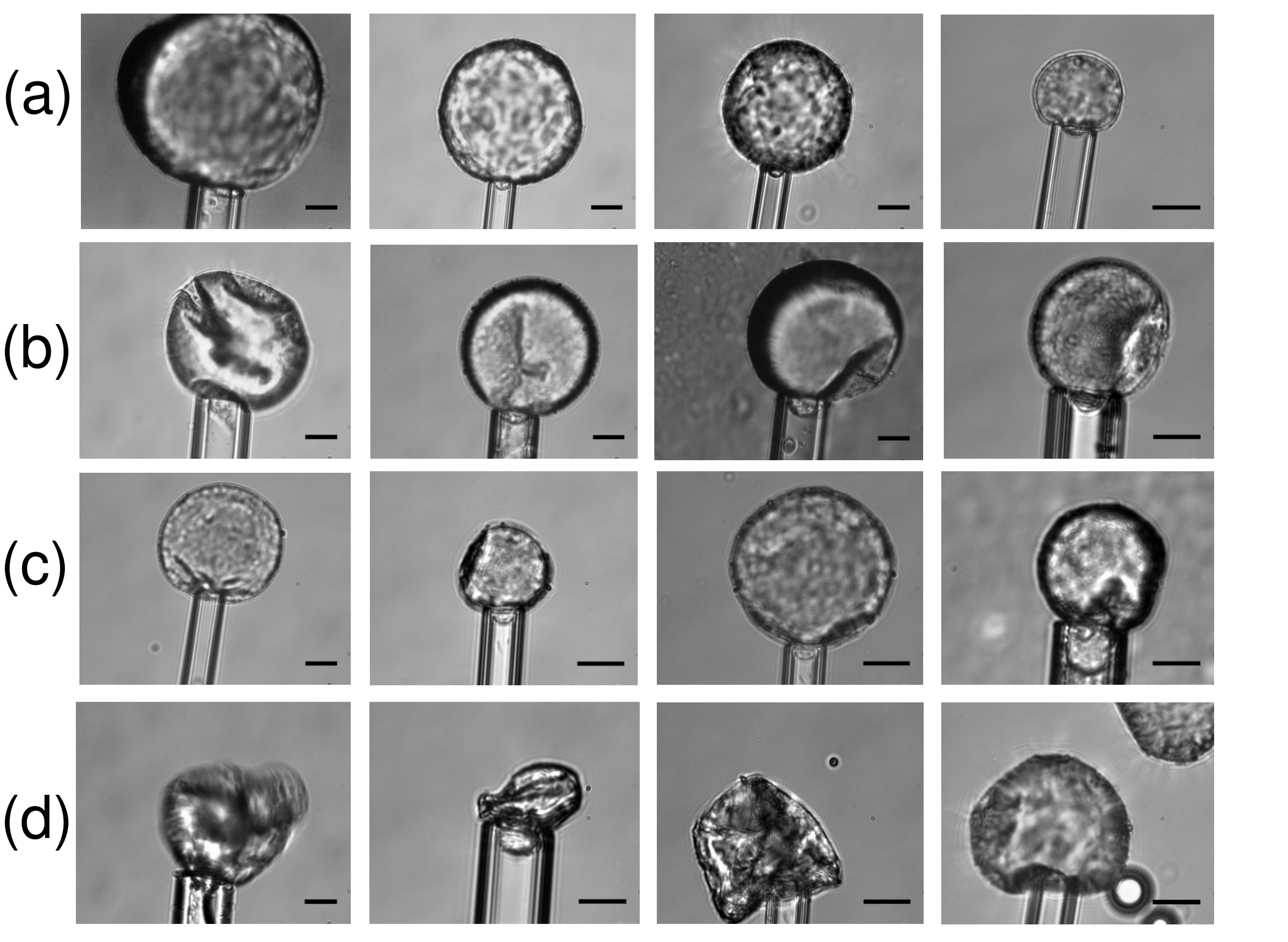}
\caption{
\textit{Gallery of elastic instabilities.} 
At the onset of buckling, we observe surface deformations including \textbf{(a)} multiple dimples, \textbf{(b)} large surface folds, \textbf{(c)} localized wrinkles, and \textbf{(d)} catastrophic failure.
The scale bar in all images is 20 $\mu$m.  
}
\label{Fig2} 
\end{figure}

As the droplet shrinks with further suction, it eventually loses structural symmetry (\textit{Supplemental Movie 2}). 
The morphology of the deformed surface is complex and highly variable.
It can develop multiple dimples (Figure \ref{Fig2} (a)),  large  surface folds (Figure \ref{Fig2} (b)), or wrinkles (Figure \ref{Fig2} (c)).
Sometimes, we  observe a sudden, catastrophic failure of the droplet (Figure \ref{Fig2} (d)).
Upon losing droplet volume or further increase in suction pressure, the surface features grow, eventually leading to the complete collapse of the droplet.
We define the suction pressure at this instability point as $\Delta P^{buc}_c$ (red line in Figure \ref{Fig1}(b)).
The variations in the buckling morphology observed likely reflect heterogeneity in the packing of particles on the droplet surface.

The droplet can fully recover its shape upon removing the suction pressure (\textit{Supplemental Movie 4}).
Upon repeated suction, the shell wrinkles at the same spots (\textit{Supplemental Movie 4}). 
This observation suggests that there are limited  particle rearrangements during the course of the droplet deformation, and that  the buckling morphology is determined by structural defects in the particle packing. 

\begin{figure}[h!]
\includegraphics[width= 0.5 \textwidth]{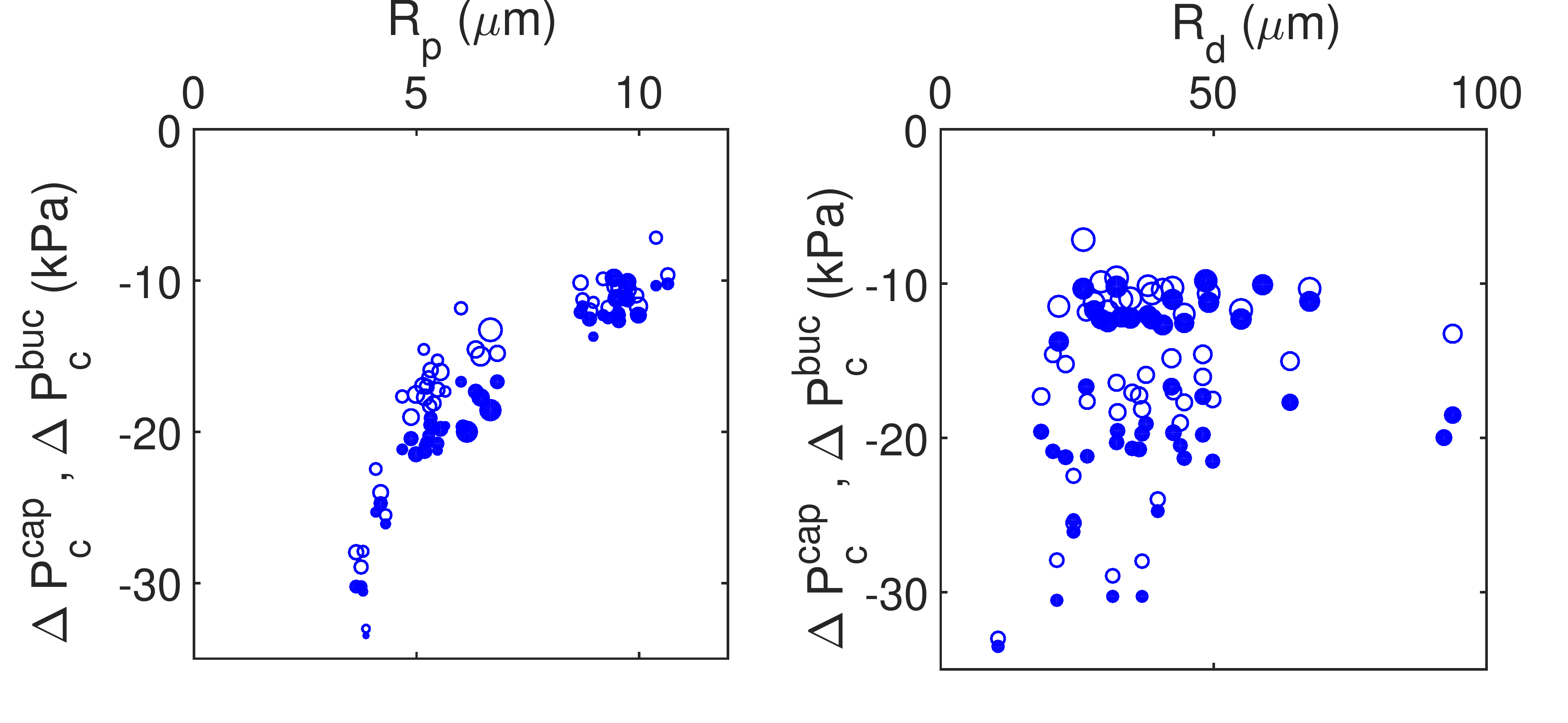}
\caption{
\textit{Critical pressure data for particle-laden droplets, varying pipette and droplet radii.} 
(\textbf{Left}) Critical pressures, $\Delta P^{cap}_{c}$  (open circles) and $\Delta P^{buc}_{c}$ (filled circles) versus pipette radius $R_p$.
Marker size scales with corresponding droplet radius $R_d$.
(\textbf{Right}) The same critical pressures versus $R_d$.
Marker size now scales with corresponding $R_p$.
}
\label{Fig3} 
\end{figure}

We have seen that particle-laden droplets fail under suction in two steps: first, a capillary instability, where encapsulated fluid escapes the droplet, and second, an elastic instability, where the shell buckles. 
To reveal the underlying physics, we measure the onset of these two instabilities as a function of the droplet radius, $R_d$, and pipette radius, $R_p$.
We vary $R_d$ from 10 - 100 $\mu$m and $R_p$ from 3 - 12 $\mu$m. 
Figure \ref{Fig3} reports $\Delta P^{cap}_{c}$ with open circles and $\Delta P^{buc}_{c}$ with filled circles for the above range of $R_d$ and $R_p$.
Here, the marker size in the left panel scales with the droplet radius and the marker size in the right panel scales with the pipette radius.  
While the capillary instability always precedes buckling, $\Delta P^{buc}_{c}\leqslant \Delta P^{cap}_{c}$, both instabilities occur at a similar magnitude of suction, $|\Delta P^{buc}_{c}-\Delta P^{cap}_{c}| \ll |\Delta P^{cap}_{c}|$.
The magnitude of the critical pressures decreases strongly with the pipette radius; however, they do not show strong dependence on the droplet size.

To understand how this size dependence reflects the mechanical properties of the interface, let us  first review  the case of a bare droplet.
The interface of a bare fluid droplet has a simple state of stress: its tension is independent of shape or deformation, $\tau=\gamma$.
Mechanical equilibrium of fluid interfaces is well-captured by the Young- Laplace equation, which states that $\Delta P = \tau \kappa$, where $\tau$ is the surface stress and $\kappa$ is the total interfacial curvature, $2/R$ for a sphere of radius $R$.
Therefore, as the pressure in the pipette is reduced, the curvature of the tongue must increase.
However, the curvature of the tongue is limited by the radius of the pipette.
Thus, at the limit of mechanical stability \cite{studMPApaper},
\begin{equation}
\label{Laplace}
\Delta P= P^{(pip)}-P^{(atm)}=2 \tau \left[ \frac{1}{R_d} - \frac{1}{R_p} \right].
\end{equation}

\noindent or in terms of the tension,
\begin{equation}
\label{tension}
\tau=\frac{\Delta P}{2 (1/R_d - 1/R_p)}. 
\end{equation}

\begin{figure}
\includegraphics[width=0.5 \textwidth]{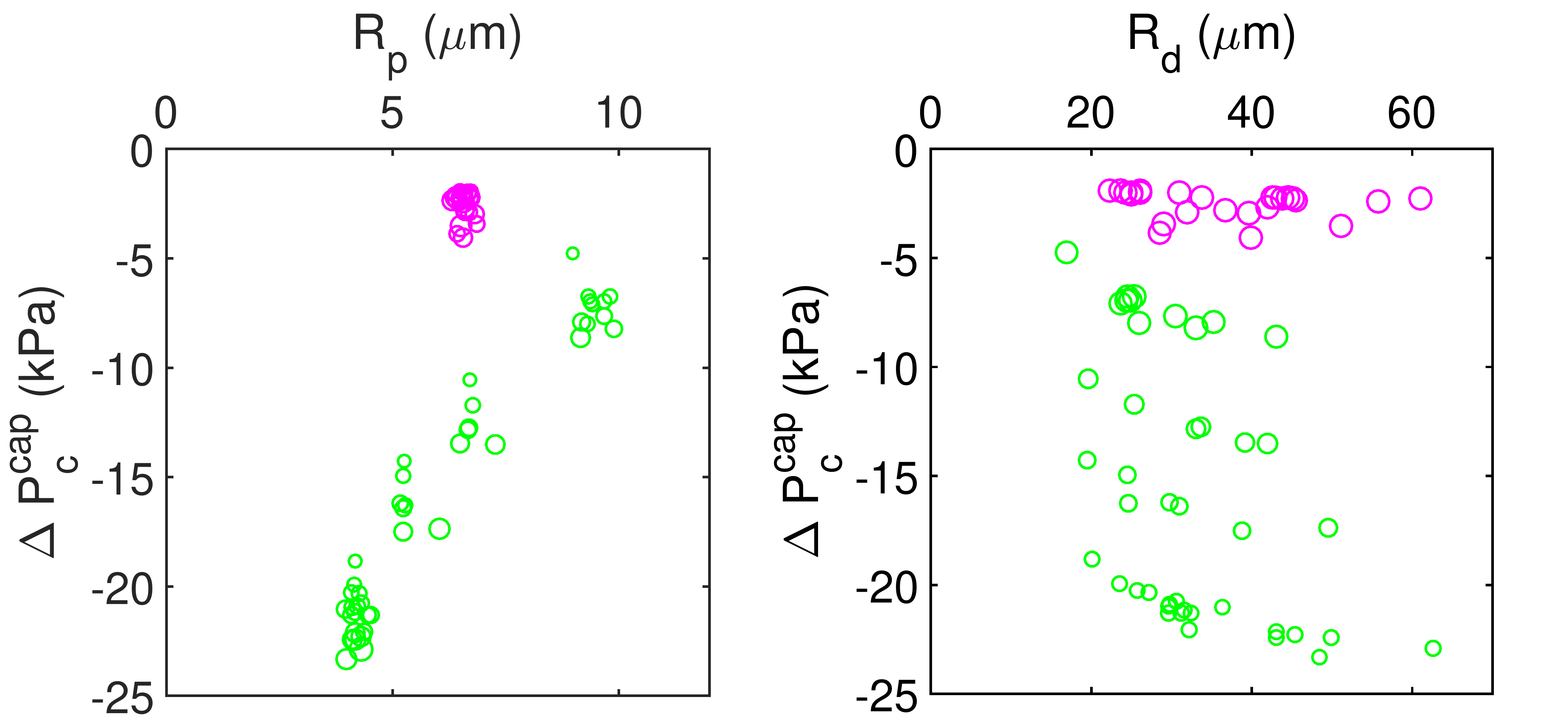}
\caption{
\textit{Critical pressure $\Delta P^{cap}_c$ for bare droplets (green) and bare droplets in 1$\%$ SDS (magenta).} 
(\textbf{Left}) Critical pressure $\Delta P^{cap}_{c}$  versus pipette radius $R_p$.
Marker size scales with corresponding droplet radius $R_d$.
(\textbf{Right}) The same $\Delta P^{cap}_{c}$ versus $R_d$.
Marker size now scales with corresponding $R_p$.  
}
\label{Fig4} 
\end{figure}

\begin{figure}
\includegraphics[width= 0.5 \textwidth]{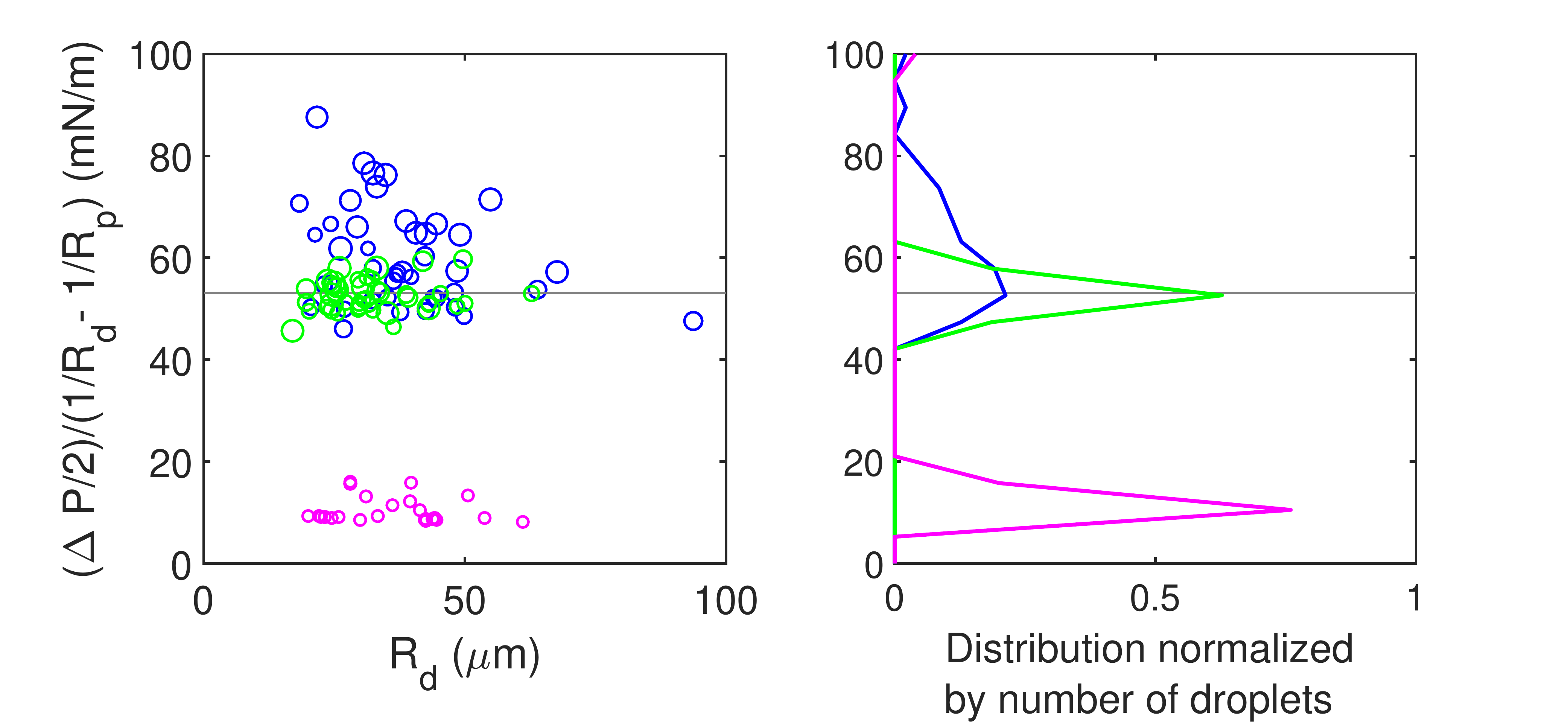}
\caption{
\textit{Apparent tension at the onset of the capillary instability,  $\tau_{c}$, based on Laplace equation (\ref{tension}) for various droplets} 
\textbf{(Left)}  $\tau_{c}$ versus droplet radius $R_d$ for particle-laden droplets (blue), bare droplets (green), and  bare droplets in 1\% SDS (magenta). 
Solid line is the expected surface tension for a bare droplet, $\gamma = 53.1$ mN/m \cite{Baretension}.
\textbf{(Right)} Spread in $\tau_c$ normalized by number of droplets.   
}
\label{Fig5}
\end{figure}

We measured the onset of the capillary instability for a range of bare hexadecane droplets in water.
The critical pressure for these droplets shows a clear dependence on the droplet and pipette size  (Figure \ref{Fig4}, green).
However, when we use  Equation \ref{tension} and the droplet radius $R_d$ at the onset of capillary instability to calculate the tension, we find that it is independent of $R_d$ or $R_p$ and is $\tau_{c} = 52.5 \pm 3.1$ mN/m  (Figure \ref{Fig5}, green).
This is very close to the surface tension of hexadecane-water interface reported in the literature, $\gamma =53.1$ mN/m \cite{Baretension}.
Therefore, the marginal mechanical stability of a bare liquid interface is well described by Equations \ref{Laplace} and \ref{tension} with a constant tension.

For comparison, we prepared hexadecane droplets stabilized in a solution of  1\% (by weight) SDS (American Bioanalytical ABO1920-00100), a common molecular surfactant.
Upon aspiration, these droplets respond similarly to  bare droplets.  
They undergo a single catastrophic capillary instability without any visible buckling (Figure \ref{Fig4}, magenta).
Like the bare droplet case, we find that the mechanical stability of the droplets is determined by  Equations \ref{Laplace} and \ref{tension} with a size-independent tension $\tau_c =10.2 \pm 2.5$ mN/m (Figure \ref{Fig5}, magenta).
The tension here is much lower than the bare interface case, and  is consistent with literature values of the water/SDS/hexadecane surface tension  \cite{bareSDS}.
So although SDS stabilizes the interface, it makes the droplets mechanically weaker and hence, easier to deform.

Given its success in describing the bare and SDS-laden droplets, we apply Equation \ref{tension} to calculate the critical tension for the capillary instability of the particle-laden droplets and plot these results in Figure \ref{Fig5}; blue.
While the largest droplets appear to have a tension similar to the bare droplets, the apparent tension of the smaller droplets is more variable, and greater than or equal to the tension of the bare droplets.
This size dependence and the scatter suggests that Equation \ref{Laplace} is not appropriate to describe the marginal equilibrium of particle-stabilized droplets, and that the tension values that Equation \ref{tension} returns are incorrect. 

Why do Equations \ref{Laplace} and \ref{tension} succeed for bare and SDS-laden droplets while failing for particle-laden droplets?
The essential difference lies in the rheology of the interface.  
While the bare and SDS-laden interfaces are fluid \cite{rheologySDS}, the particle-laden interface is solid.
This is a generic feature of particle-laden interfaces at sufficiently high density \cite{rheologyreview} \cite{JanGerry} and is made apparent in this system by the lack of particle rearrangements at the interface, as described above.
For a solid interface, the tension is no longer uniform across the interface, and the state of stress is more complex than assumed in the arguments leading up to Equation \ref{Laplace}.
 
We make progress by noting that while aspiration compresses the interface  outside the pipette, it dilates the interface inside the pipette.  
The interface of the tongue should therefore have liquid-like rheology with a tension nearly equal to that of the bare interface. 
Hence, we can determine the pressure inside the droplet at the point of instability using the Young-Laplace equation, $\Delta P^{drop}_{c} = \Delta P_{c}+2 \gamma_{hex-water}/R_p$.

\begin{figure}
\begin{center}
\includegraphics[width = 80 mm]{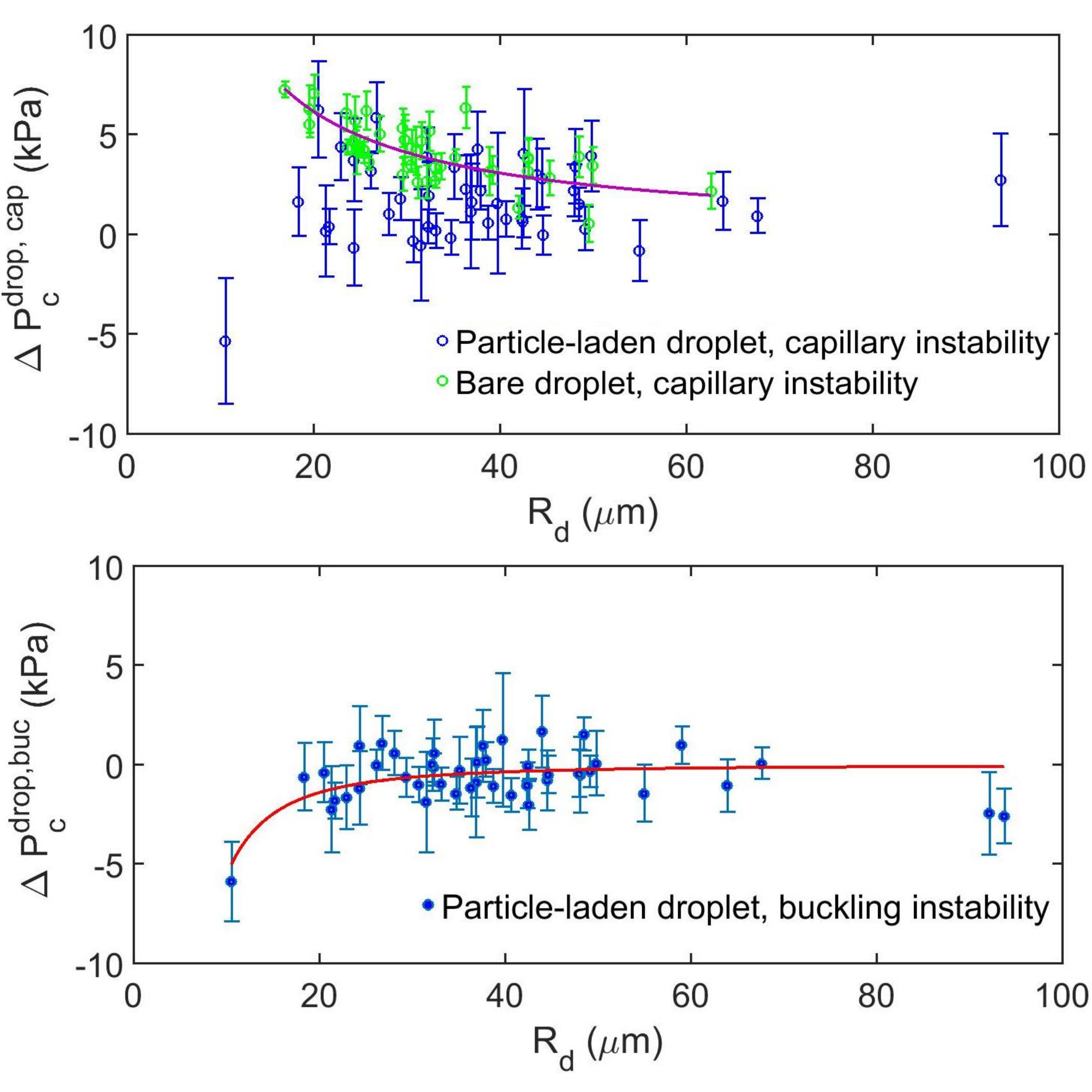}
\caption{
\textit{Pressure inside various droplets at the onset of instability} 
\textbf{(top)} $\Delta P^{drop, cap}_{c}$ at the onset of capillary instability versus droplet radius $R_d$ for particle-laden droplets (blue) and bare droplets (green).
The purple line is a Laplace fit and gives $\gamma_{fit} = 58 \pm 1 $ mN/m. 
\textbf{(bottom)} $\Delta P^{drop, buc}_{c}$ at the onset of buckling instability. 
The red curve is fit to Equation \ref{bucP} and gives $E'.t^{2} = 560 \pm 1$ kPa $\cdot\mu$m$^2$.
The error bars in the data show the standard deviation of uncertainty. 
}
\label{Fig6}
\end{center}
\end{figure}

The size dependence of the pressure inside the droplet at the onset of capillary instability, $\Delta P^{drop,cap}_{c}$, is plotted in Figure \ref{Fig6} (top).
For bare droplets, shown in green, this pressure increases as the droplet becomes smaller, as predicted by the Young-Laplace equation.
However, the pressure inside the particle-laden droplets at the point of capillary instability shows no apparent size dependence and is near zero.
The results are very similar if we consider the pressure inside the droplet at the point of buckling,  $\Delta P^{drop,buc}_{c}$, as shown in Figure \ref{Fig6} (bottom).
Again, we see very little dependence of the pressure on the droplet size, and the values clustered around zero.

The buckling of the particle-laden droplets at zero pressure is consistent with classic observations of surface instabilities of compressed monolayers \cite{Milner,Aveyard,FullerPCCP}.  
There, it is found that the surface goes unstable when the net tension on the interface approaches zero.  
The Young-Laplace equation demands that the pressure drop across an interface goes to zero as its tension goes to zero.
By that argument, the surface of particle-laden droplets should become unstable when the pressure drop across the interface vanishes.
This is consistent with the observed onset of buckling shown in Figure \ref{Fig6} (bottom).

Generally, the vanishing of the tension is insufficient to drive the surface instability since a surface can also have some bending rigidity.
The finite rigidity case is captured by the classic linear theory of thin shells \cite{landau}\cite{vonK}.
Here, a thin homogeneous linear elastic shell becomes unstable when
\begin{equation}
\label{bucP}
\Delta P^{drop,buc}_{c} = -{E'}\left(\frac{t}{R_d}\right)^2 ,
\end{equation}
where $t$ is the shell thickness and $E'$ is a modified Young's modulus, $E' =  \frac {E}{\sqrt[]{3 (1- \nu^2 )}} $.
While the particle shells are inhomogeneous, we use Equation \ref{bucP} to put an upper bound on the apparent value of $E't^2$. 
We find $E't^2$ to be $560$~ kPa$\cdot\mu$m$^2$ based on the fit  shown by the red curve in Figure \ref{Fig6} (bottom). 

\section{IV. Conclusions}

The mechanics of particle-stabilized droplets was found to be qualitatively different than the mechanics of bare or SDS-stabilized droplets.
Particle-stabilized droplets undergo a two-step failure under suction.
First, a capillary instability, where fluid is withdrawn from the shell with minimal change to the droplet shape.  Second, an elastic instability, where the shell buckles.
Both instabilities occur as the interfacial tension vanishes.
The gap in suction pressure between the two instabilities may arise from a finite, but small, bending rigidity of the particle-laden interface.
Particle-laden emulsion droplets may be an interesting application of recent advances in our understanding of the far-from-threshold deformation of highly bendable sheets \cite{BennyPNAS,VellaTensiometry,KingPNAS}.

\section{Acknowledgments}

We thank Dominic Vella, Siddharth Prabhu, and Kate Jensen for helpful discussions. 
This work was supported by a gift from AMOREPACIFIC Co. and NSF CBET 1236086.

%

\end{document}